\documentstyle[12pt,twoside,fleqn,epsf,espcrc1]{article}

\def\gsim{\mathrel{\raise.3ex\hbox{$>$}\mkern-14mu
             \lower0.6ex\hbox{$\sim$}}}
\def\lsim{\mathrel{\raise.3ex\hbox{$<$}\mkern-14mu
             \lower0.6ex\hbox{$\sim$}}}

\newcommand{\AmS}{{\protect\the\textfont2
  A\kern-.1667em\lower.5ex\hbox{M}\kern-.125emS}}

\title{What QCD Tells Us About Nature -- and Why We Should Listen
\thanks{Keynote talk at PANIC `99, Uppsala Sweden, June 10, 1999.
A Turn of the Millennium Conference. IASSNS-HEP/99-64}}

\author{Frank Wilczek\address{School of Natural Science, 
        Institute for Advanced Study\\ 
        Princeton, NJ 08540 USA}
        \thanks{Research supported in part by DOE grant
                DE-FG02-90ER40542. e-mail: wilczek@sns.ias.edu}}
       
\begin{document}

\maketitle

\begin{abstract}
I discuss why QCD is our most perfect physical theory.  Then I visit a
few of its current frontiers.  Finally I draw some appropriate
conclusions.
\end{abstract}

\section{QCD is our most perfect physical theory}

Here's why:

\subsection{It embodies deep and beautiful principles.}

These are, first of all, the general principles of quantum mechanics,
special relativity, and locality, that lead one to relativistic
quantum field theory \cite{one}.  In addition, we require invariance under the
nonabelian gauge symmetry $SU(3)$, the specific matter content of
quarks -- six spin 1/2 Dirac fermions which are color triplets -- and
renormalizability.  These requirements determine the theory completely,
up to a very small number of continuous parameters as discussed below.

Deeper consideration reduces the axioms further.  Theoretical
physicists have learned the hard way that consistent, non-trivial
relativistic quantum field theories are difficult to construct, due to
the infinite number of degrees of freedom (per unit volume) needed to
construct local fields, which tends to bring in ultraviolet
divergences.  To construct a relativistic quantum theory, one
typically introduces at intermediate stages a cutoff, which spoils the
locality or relativistic invariance of the theory.  Then one attempts
to remove the cutoff, while adjusting the defining parameters, to
achieve a finite, cutoff-independent limiting theory.  Renormalizable
theories are those for which this can be done, order by order in a
perturbation expansion around free field theory.  This formulation,
while convenient for mathematical analysis, obviously begs the
question whether the perturbation theory converges (and in practice it
never does).

A more straightforward procedure, conceptually, is to regulate the
theory as a whole by discretizing it, approximating space-time by a
lattice \cite{two}.  This spoils the continuous space-time symmetries of the
theory.  Then one attempts to remove dependence on the discretization
by refining it, while if necessary adjusting the defining parameters,
to achieve a finite limiting theory that does not depend on the
discretization, and does respect the space-time symmetries.  The
redefinition of parameters is necessary, because in refining the
discretization one is introducing new degrees of freedom.  The
earlier, coarser theory results from integrating out these degrees of
freedom, and if it is to represent the same physics it must
incorporate their effects, for example in vacuum polarization.

In this procedure, the big question is whether the limit exists.  It
will do so only if the effects of integrating out the additional
short-wavelength modes, that are introduced with each refinement of
the lattice, can be captured accurately by a re-definition of
parameters already appearing in the theory.  This, in turn, will occur
only in a straightforward way if these modes are weakly
coupled. (Another simple possibility is that short-distance modes
of different types cancel in vacuum polarization.  This is what occurs
in supersymmetric theories.  Other types of ultraviolet fixed
points are in principle possible, but difficult to imagine or
investigate.)   But this is true, if and only if the theory is
asymptotically free.

One can investigate this question, i.e., whether the couplings
decrease to zero with distance, or in other words whether the theory
is asymptotically free, within weak coupling perturbation theory
\cite{three}.  One finds that only nonabelian gauge theories with
simple matter content, and no non-renormalizable couplings, satisfy
this criterion \cite{four}.  Supersymmetric versions of these theories
allow more elaborate, but still highly constrained, matter content.

Summarizing the argument, only those relativistic field theories which
are asymptotically free can be argued in a straightforward way to
exist.  And the only asymptotically free theories in four space-time
dimensions involve nonabelian gauge symmetry, with highly restricted
matter content.  So the axioms of gauge symmetry and renormalizability
are, in a sense, gratuitous.  They are implicit in the mere {\it
existence\/} of non-trivial interacting quantum field theories.

Thus QCD is a member of a small aristocracy: the closed, consistent
embodiments of relativity, quantum mechanics, and locality.  Within
this class, it is among the least affected members.

\subsection{It provides algorithms to answer any physically meaningful question within its scope.}

As I just discussed, QCD can be constructed by an explicit, precisely
defined discretization and limiting procedure.  This provides, in
principle, a method to compute any observable, in terms that could be
communicated to a Turing machine.

In fact marvelous things can be accomplished, in favorable cases,
using this direct method.  For a stirring example, see Figure 4 below.

The computational burden of the direct approach is, however, heavy at
best.  When one cannot use importance sampling, as in addressing such
basic questions as calculating scattering amplitudes or finding the
ground state energy at finite baryon number density, it becomes
totally impractical.

For these reasons various improved perturbation theories continue to
play an enormous role in our understanding and use of QCD.  The most
important and well-developed of these, directly based on asymptotic
freedom, applies to hard processes and processes involving heavy
quarks \cite{five}.  It is what is usually called ``perturbative QCD'', and
leads to extremely impressive results as exemplified by Figures 1-3,
below.  The scope of these methods is continually expanding, to
include additional ``semi-hard'' processes, as will be discussed
in many talks at the Conference.  When combined with some additional
ideas, they allow us to address major questions regarding the behavior
of the theory at high temperatures and large densities, as I'll
touch on below.

Chiral perturbation theory \cite{six}, which is based on quite
different aspects of QCD, is extremely useful in discussing low-energy
processes, though it is difficult to improve systematically.  A proper
discussion of this, or of many other approaches each of which offers
some significant insight (traditional nuclear physics, bag model,
Reggeism, sum rules, large N, Skyrme model, ... ) would not be
appropriate here.

I would however like to mention a perturbation theory which I think is
considerably underrated, that is strong coupling perturbation theory
\cite{seven}.  It leads to a simple, appealing, and correct
understanding of confinement \cite{eight}, and even its existing, crude
implementations provide a remarkably good caricature of the low-lying
hadron spectrum.  It may be time to revisit this approach, using
modern algorithms and computer resources.

\subsection{Its scope is wide.}

There are significant applications of QCD to nuclear physics,
accelerator physics, cosmology, extreme astrophysics, unification, and
natural philosophy.  I'll say just a few words about each, in turn.

{\it Nuclear physics} : Understanding atomic nuclei was of course the
original goal of strong interaction physics.  In principle QCD
provides answers to all its questions.  But in practice QCD has not
superseded traditional nuclear physics within its customary domain.  The
relationship between QCD and traditional nuclear physics is in some
respects similar to the relationship between QED and chemistry.  The
older disciplines retain their integrity and independence, because
they tackle questions that are exceedingly refined from the point of
view of the microscopic theories, involving delicate cancellations and
competitions that manufacture small net energy scales out of much
larger gross ones.  QCD offers many insights and suggestions, however,
of which we will hear much at this Conference.  There is also an
emerging field of extreme nuclear physics, including the study of
nuclei with hard probes and heavy-ion collisions, where the influence
of QCD is decisive.

{\it Accelerator physics} : Most of what goes on at high-energy
accelerators is described by QCD.  This application has been so
successful, that experimenters no longer speak of ``tests of
QCD'', but of ``QCD backgrounds''!  Two- and even three-loop
calculations of such ``backgrounds'' are in urgent demand.  What
can one add to that sincere testimonial?

{\it Cosmology} : Because of asymptotic freedom, hadronic matter
becomes not impenetrably complex, but rather profoundly simpler, under
the extreme conditions predicted for the early moments of the Big
Bang.  This stunning simplification has opened up a large and fruitful
area of investigation.

{\it Extreme astrophysics} : The physics of neutron star interiors,
neutron star collisions, and collapse of very massive stars involves
extreme nuclear physics.  It should be, and I believe that in the
foreseeable future it will be, firmly based on microscopic QCD.

{\it Unification and Natural philosophy} : See below.

\subsection{It contains a wealth of phenomena.}

Let me enumerate some major ones: radiative corrections, running
couplings, confinement, spontaneous (chiral) symmetry breaking,
anomalies, instantons.  Much could be said about each of these, but I
will just add a few words about the first.  The Lamb shift in QED is
rightly celebrated as a triumph of quantum field theory, because it
shows quantitatively, and beyond reasonable doubt, that loop effects
of virtual particles are described by the precise, intricate rules of
that discipline.  But in QCD, we probably have by now 50 or so cases
where two- and even three-loop effects are needed to do justice to
experimental results -- and the rules are considerably more intricate!

\subsection{It has few parameters ... }

A straightforward accounting of the parameters in QCD would suggest 8:
the masses of six quarks, the value of the strong coupling, and the
value of the P and T violating $\theta$ parameter.  The fact that
there are only a small finite number of parameters is quite profound.
It is a consequence of the constraints of gauge invariance and
renormalizability (or alternatively, as we saw, existence, by way of
asymptotic freedom).

In reality there are not 8 parameters, but only 6.  $g$ is eliminated
by dimensional transmutation \cite{nine}.  This means, roughly stated, that
because the coupling runs as a function of distance, one cannot
specify a unique numerical value for it.  It will take any value, at some
distance or other.  One can put (say) $g(l) \equiv 1$, thereby
determining a length scale $l$.  What appeared to be a choice of
dimensionless coupling, is revealed instead to be a choice of unit of
length, or equivalently (with $\hbar = c = 1$) of mass.  Only the
dimensionless {\it ratio\/} of this mass to quark masses can enter
into predictions for dimensionless quantities.  So what appeared to be
a one-parameter family of theories, with different couplings, turns
out to be a single theory measured using differently calibrated
meter-sticks.

The $\theta$ term is eliminated, presumably, by the Peccei-Quinn
mechanism \cite{ten}.  To assure us of this, it would be very nice to
observe the quanta associated with this mechanism, namely axions
\cite{eleven}.  In any case, we know for sure that the $\theta$ term is
very small.  For purposes of strong interaction physics, within QCD
itself, we can safely set it to zero, invoking P or T symmetry, and be
done with it.

\subsection{... or none.}

This economy of parameters would already be quite impressive, given
the wealth of phenomena described.  However if we left it at that we
would be doing a gross injustice to QCD, and missing one of its most
striking features.

To make my point, let me call your attention to a simplified version
of QCD, that I call ``QCD Lite''.  QCD Lite is simply QCD
truncated to contain just two flavors of quarks, both of which are
strictly massless, and with the $\theta$ parameter set to zero.  These
choices are natural in the technical sense, since they can be replaced
by symmetry postulates.  Indeed, assuming masslessness of the quarks
is tantamount to assuming exact $SU(2)\times SU(2)$ chiral symmetry,
and $\theta = 0$ is tantamount to assuming the discrete symmetries P
or T.  (Actually, once we have set the quark masses to zero, we can
dial away $\theta$ by a field redefinition.)

Now there are two especially remarkable things about QCD Lite \cite{twelve}.  The
first is that it is a theory which {\it contains no continuous free
parameters at all}.  Its only inputs are the numbers 3 (colors) and 2
(flavors).  The second is that it {\it provides an excellent
semi-quantitative theory of hadronic matter}.

Indeed, in reality the strange and heavier quarks have very little
influence on the structure or masses of protons, neutrons, atomic
nuclei, pions, rho mesons, ... .  Leaving them out would require us to
abridge, but not to radically revise, the Rosenfeld Tables.  This is
proved by the remarkable quantitative success, at the 5-10\% level, of
lattice gauge theory in the `quenched' approximation \cite{thirteen}.
For in this approximation, the influence of the heavier quarks on the
lighter ones is systematically ignored.

The only major effect of putting the u and d masses to zero is to make
the pions, which are already quite light by hadronic standards,
strictly massless.  The perturbation to the proton mass, for instance,
can be related using chiral perturbation theory to the value of the so-called
$\sigma$ term, a directly measurable quantity \cite{fourteen}.  When
this is done, one finds that the u and d quark masses are responsible
for only about 5\% of the proton mass.

Thus QCD Lite provides a truly remarkable realization of John
Wheeler's program, ``Getting Its From Bits''.  For here we encounter
an extremely rich and complex class of physical phenomena --
including, in principle, nuclear and particle spectra -- that can be
calculated, accurately and without ambiguity, using as sole inputs the
numbers 3 and 2.

\subsection{It is true.}
I will not waste a lot of words on this, showing instead a few pictures,
each worth many thousands of words.

Figure 1 \cite{fifteen} displays graphically that many independent types of
experiments at different energy scales have yielded determinations of
the strong coupling constant, all consistent with the predicted
running.  The overall accuracy and consistency of this phenomenology
is reflected in the precision with which this coupling is determined,
to wit 5\% (at the Z mass).  A remarkable feature of the theory is
that a wide range of possible values for the coupling at relatively
small energies focuses down to quite a narrow range at the highest
accessible energies.  Thus any ``reasonable'' choice of the scale
at which the coupling becomes numerically large leads, within a few
per cent, to a unique value of the coupling at the Z mass.  Our
successful QCD predictions for high energy experiments have no
wiggle-room!

While Figure 1 is impressive, it does not do complete justice to the
situation.  For several of the experimental `points' each represents a
summary of hundreds of independent measurements, any one of which
might have invalidated the theory, and which display many interesting
features.  Figures 2 and 3 partially ameliorate the omission.  Figure
2 \cite{sixteen} shows some of the experimental data on deep inelastic
scattering -- all subsumed within the `DIS' point in Figure 1 --
unfolded to show the complete $Q^2$ and $x$ dependence.  Our
predictions \cite{five} for the pattern of evolution of structure
functions with $Q^2$ --decrease at large $x$, increase at small $x$
-- are now confirmed in great detail, and with considerable precision.
Particularly spectacular is the rapid growth at small $x$.  This was
predicted \cite{seventeen}, in the form now observed, very early on.
However, even at the time we realized this rise could not continue 
forever.  The proliferating partons begin to form a dense system, and
eventually one must cease to regard them as independent.  There is a
very interesting many-body problem developing here, which seems ripe
for experimental and theoretical investigation, and may finally allow
us to make contact between microscopic QCD and the remarkably
successful Regge-pole phenomenology.

Figure 3 \cite{eighteen} displays a comparison of the experimental
distribution of jet energies, in 3-jet events, with the QCD
prediction.  Shown is the energy fraction of the second hardest jet,
compared to its kinematic maximum.  For a detailed explanation, see
\cite{eighteen}.  The rise at $x_2 \rightarrow 1$ reflects the
singularity of soft gluon bremsstrahlung, and matches the prediction
of QCD (solid line).  For comparison, the predictions for hypothetical
scalar gluons are shown by the dotted line.  This is as close to a
direct measurement of the core interaction of QCD, the basic
quark-gluon vertex, as you could hope to see.  The other piece of this
Figure displays a related but more sophisticated comparison, using the
Ellis-Karliner angle.

Finally Figure 4 \cite{nineteen} shows the comparison of the QCD
predictions to the spectrum of low-lying hadrons.  Unlike what was
shown in the previous two Figures, and most of the points in Figure 1,
this tests the whole structure of the theory, not only its
perturbative aspect.  The quality of the fit is remarkable.  Note that
only one adjustable parameter (the strange quark mass) and one overall
choice of normalization go into the calculation.  Otherwise it's pure
``Its From Bits''.  Improvements due to enhanced computing power and
to the use of domain wall quarks \cite{twenty}, that more nearly respect chiral
symmetry, are on the horizon.

Since there seems to be much confusion (and obfuscation) on the point,
let me emphasize an aspect of Figure 4 that ought to be blindingly
clear: what you don't see in it.  You don't see massless degrees
of freedom with long-range gauge interactions, nor parity doublets.
That is, confinement and chiral symmetry breaking are simply true
facts about the solution of QCD, that emerge by direct calculation.
The numerical work has taken us way beyond abstract discussion of
these features.

\subsection{It lacks flaws.}

Finally, to justify the adverb in ``{\it most\/} perfect'' I
must briefly recall for you some prominent flaws in our other best
theories of physics, which QCD does not share.

Quantum electrodynamics is of course extremely useful -- incomparably
more useful than QCD -- and successful in practice.  But there is a
worm in its bud. It is not asymptotically free.  Treated outside of
perturbation theory, or extrapolated to extremely high energy, QED
becomes internally inconsistent.  Modern electroweak theory shares
many of the virtues of QED, but it harbors the same worm, and in
addition contains many loose ends and continuous free parameters.
General relativity is the deepest and most beautiful theory of all,
but it breaks down in several known circumstances, producing
singularities that have no meaningful interpretation within the
theory.  Nor does it mesh seamlessly with the considerably better
tested and established framework we use for understanding the
remainder of physics.  Specifically, general relativity is notoriously
difficult to quantize.  Finally, it begs the question of why the
cosmological term is zero, or at least fantastically small when
measured in its natural units.  Superstring theory promises to solve
some and conceivably all of these difficulties, and to provide a fully
integrated theory of Nature, but I think it is fair to say that in its
present form superstring theory is not defined by clear principles,
nor does it provide definite algorithms to answer questions within its
claimed scope, so there remains a big gap between promise and
delivery.

\section{Breaking New Ground in QCD, 1: High Temperature}

The behavior of QCD at high temperature, and low baryon number
density, is relevant to cosmology -- indeed, it describes the bulk of
matter filling the Universe, during the first few seconds of the Big
Bang -- and to the description of both numerical and physical
experiments.  There are ambitious experimental programs
planned for RHIC, and eventually LHC, to probe this physics.  It is
also, I think, intrinsically fascinating to ask -- what happens to
empty space, if you keep adding heat?

The equilibrium thermodynamics of QCD at finite temperature (and zero
chemical potential) is amenable to direct simulation, using the
techniques of lattice gauge theory.  Figure 5 \cite{twentyone} does
not quite represent the current, rapidly evolving, state of the art,
but it does already demonstrate some major qualitative points.

The chiral symmetry breaking condensate, clearly present (as
previously advertised) at zero temperature, weakens and seems to be
gone by $T \sim 150 ~ {\rm Mev}$.  Likewise at these temperatures there
is a sizable increase in the value of the Polyakov loop, indicating
that the force between distant color sources has considerably
weakened.  Furthermore the energy density increases rapidly,
approaching the value one would calculate for an ideal gas of quarks
and gluons.  The pressure likewise increases rapidly, but lags
somewhat behind the ideal gas value.

All these phenomena indicate that at these temperatures and above a
description using quarks and gluons as the degrees of freedom is much
simpler and more appropriate than a description involving ordinary
hadrons.  Indeed, the quarks and gluons appear to be quasi-free.  That
is what one expects, from asymptotic freedom, for the high-energy
modes that dominate the thermodynamics.

It is an interesting challenge to reproduce the pressure analytically
\cite{twentytwo}.  Since the only scale in the problem is the large
temperature, if one can organize the calculation so as to avoid
infrared divergences, asymptotic freedom will legitimize a weak
coupling treatment.  Even more interesting would be to do this by a
method that also works at finite density, since the equation of state
is of great interest for astrophysics and is {\it not\/} accessible
numerically.

There is no doubt, in any case, that QCD predicts the existence of a
quark-gluon plasma phase, wherein its basic degrees of freedom,
normally hidden, come to occupy center stage.

While transition to a quark-gluon plasma at asymptotically high
temperatures is not unexpected, the abruptness and especially the {\it
precocity\/} of the change is startling.  Below 150 Mev the only
important hadronic degrees of freedom are the pions.  Why does this
rather dilute pion gas suddenly go berserk?

The change is enormous, quantitatively.  The pions represent precisely
3 degrees of freedom.  The free quarks and gluons, with all their
colors, spins, and antiparticles, represent 52 degrees of freedom.

There are many ideas for detecting signals of quark-gluon plasma
formation in heavy ion collisions, which you will be hearing much of.
I would like discuss briefly a related but more focused question, on
which there has been dramatic progress recently \cite{twentythree}.
This is the question, whether there is a true phase transition in QCD
accessible to experiment.

One might think that the answer is obviously ``yes'', since
there are striking differences between ordinary hadronic matter and
the quark-gluon plasma.  This is not decisive, however.  Let me remind
you that the dissociation of ordinary atomic gases into plasmas is not
accompanied by a phase transition, even though these states of matter
are very different (so different, that at Princeton they are studied
on separate campuses).  Similarly, confinement of quarks is believed
to go over continuously, at high temperature, into screening --
certainly, no one has demonstrated the existence of an order parameter
to distinguish between them.

What about chiral symmetry restoration?  For massless quarks, there is
a definite difference between the low-temperature phase of broken
chiral symmetry and the quasi-free phase with chiral symmetry
restored, so there must be a phase transition.  A rather subtle
analysis using the renormalization group indicated that for two
massless quarks one might have a second-order phase transition, while
for three massless quarks it must be first-order \cite{twentyfour}.
This is the pattern observed in numerical simulations.

The real world has two very light quarks and one (the strange quark)
whose mass is neither clearly small nor clearly large compared to
basic QCD scales.  Here, then, a sharp question emerges.  If the
strange quark is effectively heavy, and the other quarks are taken
strictly massless, we should have a second-order transition.  If the
strange quark is effectively light, we should have a first-order
transition.  Which is the case, for the physical value of the strange
quark mass?  Although this has been a controversial question, there
seems to be an emerging consensus among lattice gauge theorists that
it is second-order.

Unfortunately, this means that with small but finite $u$ and $d$ quark
masses we will not have a sharp phase transition at all, but only a
crossover.  And not a particularly sharp one, at that.  For while we
are ordinarily encouraged to treat these masses as small
perturbations, they are responsible for the pion masses, which are far
from negligible at the temperatures under discussion.  So the relevant
correlation length never gets very large.

Nevertheless it was interesting to point out \cite{twentyfive} that in
the $m_s -T$ plane one could naturally connect the first- and second-
order behaviors, as in Figure 6.  The line of first-order transitions
ends at a tricritical point.  This is a true critical point, with
diverging correlation lengths and large fluctuations.  The first-order
line, since it is a locus of discontinuities, and therefore the
existence of a tricritical point where it terminates, are features
which survive the small perturbation due to non-zero light quark
masses.  All these statements can be tested against numerical
simulation.

Stephanov, Rajagopal and Shuryak \cite{twentythree} have brought the
subject to a new level of interest, taking off from the simple but
brilliant observation that one expects similar behavior in the $\mu -
T$ plane.  The big advantage of this is, that while $m_s$ is not a
control parameter one can vary experimentally, the chemical potential
$\mu$ is.  They have proposed quite specific, characteristic
signatures for passage near this transition in the thermal history of
a fireball, such as might be obtained in heavy ion collisions
\cite{twentysix}.  The signatures involve enhanced fluctuations and
excess, non-thermal production of low-energy pions.

I believe it ought to be possible to refine the prediction, by
locating the tricritical point theoretically.  For while it is
notoriously difficult to deal with large chemical potentials at small
temperature numerically, there are good reasons to be optimistic about
high temperatures and relatively small chemical potentials, which is
our concern here \cite{twentyseven}.

If all these strands can be brought together, it will be a wonderful
interweaving of theory, experiment, and numerics.

\section{Breaking New Ground in QCD, 2:  High Density}

The behavior of QCD at high baryon number density, and low
temperature, is of direct interest for describing neutron star
interiors, neutron star collisions, and events near the core of
collapsing stars.  Unfortunately, it has proved quite difficult to
calculate this behavior directly numerically using lattice gauge
theory techniques.  This is because in the presence of a chemical
potential the functional integral for the partition function is no
longer positive definite (or even real) configuration by
configuration, so importance sampling fails, and the calculation
converges only very slowly.

On the other hand, there has been remarkable progress on this problem
over the last year or two using analytical techniques.  This has shed
considerable new light on many aspects of QCD.  We have new, fully
calculable mechanisms for confinement and chiral symmetry breaking
\cite{twentyeight}.  Amazingly, we find that two famous, historically
influential ``mistakes'' from the prehistory of QCD -- the Han-Nambu
\cite{twentyfour} assignment of integer charge to quarks, and the
Sakurai \cite{thirty} model of vector mesons as Yang-Mills fields
\cite{thirtyone} -- emerge from the microscopic theory at high
density.  And we find that in the slightly idealized version of QCD
with three degenerate light quarks, there need be no phase transition
separating the calculable high density phase from (the appropriate
version of) nuclear matter \cite{hadron}!

At the request of the organizers I will be giving a separate seminar
on these developments \cite{thirtytwo}, so here I will be telegraphic.

Why might we expect QCD to become analytically tractable at high
density?  At the crudest heuristic level, it is a case of asymptotic
freedom meets the fermi surface.

Let us suppose, optimistically, that a weak coupling treatment is
going to be appropriate, and see where it leads.

If the coupling is weak and the density large, our first approximation
to the ground state is large fermi balls for all the quarks.  Due to
the Pauli exclusion principle, the modes deep within the ball will be
energetically costly to excite, and the important low-energy degrees
of freedom will be the modes close to the fermi surface.  But these
modes will have large momentum.  Thus their interactions, generically,
will either hardly deflect them, or will involve large momentum
transfer.  In the first case we don't care, while the second
involves a weak coupling, due to asymptotic freedom.

On reflection, one perceives two big holes in this argument.  First,
it doesn't touch the gluons.  They remain massless, with singular
interactions and strong couplings in the infrared that do not appear
to be under control.  Second, as we learn in the theory of
superconductivity, the fermi surface is generically unstable, even at
weak coupling.  This is because pairs of particles (or holes) of equal
and opposite momenta are low-energy excitations which can all scatter
into one another.  Thus one is doing highly degenerate perturbation
theory, and in that circumstance even a small coupling can have large
qualitative effects.

Fortunately, our brethren in condensed matter physics have taught us
how to deal with the second problem, and its proper treatment also
cures the first.  There is an attractive interaction between quarks on
the opposite sides of the fermi surface, and they pair up and
condense.  In favorable cases -- and in particular, for three
degenerate or nearly degenerate flavors -- this color
superconductivity produces a gap for all the fermion excitations, and
also gives mass to all the gluons.

Thus a proper weak-coupling treatment automatically avoids all
potential infrared divergences, and our optimistic invocation of
asymptotic freedom provides, at asymptotically high density, its own
justification.

\section{Breaking New Ground in QCD, 3: Unification}

The different components of the standard model have a similar
mathematical structure, all being gauge theories.  Their common
structure encourages the speculation that they are different facets of
a more encompassing gauge symmetry, in which the different strong and
weak color charges, as well as electromagnetic charge, would all
appear on the same footing.  The multiplet structure of the quarks and
leptons in the standard model fits beautifully into small
representations of unification groups such as $SU(5)$ or $SO(10)$.
There is the apparent difficulty, however, that the coupling strengths
of the different standard model interactions are widely different,
whereas the symmetry required for unification requires that they share
a common value.

The running of couplings suggests an escape from this impasse
\cite{thirtythree}.  Since the strong, weak, and electromagnetic
couplings run at different rates, their inequality at currently
accessible scales need not reflect the ultimate state of affairs.  We
can imagine that spontaneous symmetry breaking -- a soft effect -- has
hidden the full symmetry of the unified interaction.  What is really
required is that the fundamental, bare couplings be equal, or in more
prosaic terms, that the running couplings of the different
interactions should become equal beyond some large scale.

Using simple generalizations of the formulas derived and tested in
QCD, which are none other than the ones experimentally validated in
Figure 1, we can calculate the running of all the couplings, to see
whether this requirement is met.  In doing so one must make some
hypothesis about the spectrum of virtual particles.  If there are
additional massive particles (or, better, fields) that have not yet
been observed, they will contribute significantly to the running of
couplings once the scale exceeds their mass.

Let us first consider the default assumption, that there are no new
fields beyond those that occur in the standard model.  The results of
this calculation are displayed in Figure 7 \cite{thirtyfour}.
Considering the enormity of the extrapolation this works remarkably
well, but the accurate experimental data indicates unequivocally that
something is wrong.

There is one particularly attractive way to extend the standard model,
by including supersymmetry.  Supersymmetry cannot be exact, but if it
is only mildly broken (so that the superpartners have masses $\lsim$ 1
Tev) it can help explain why radiative corrections to the Higgs mass
parameter, and thus to the scale of weak symmetry breaking, are not
enormously large.  In the absence of supersymmetry power counting
would indicate a hard, quadratic dependence of this parameter on the
cutoff.  Supersymmetry removes the most divergent contribution, by
canceling boson against fermion loops.  If the masses of the
superpartners are not too heavy, the residual finite contributions due
to supersymmetry breaking will not be too large.  The minimal
supersymmetric extension of the standard model, then, makes definite
predictions for the spectrum of virtual particles starting at 1 Tev or
so.  Since the running of couplings is logarithmic, it is not
extremely sensitive to the unknown details of the supersymmetric mass
spectrum, and we can assess the impact of supersymmetry on the
unification hypothesis quantitatively.  The results, as shown in
Figure 8 \cite{thirtyfive}, are quite encouraging.

A notable result of the unification of couplings calculation,
especially in its supersymmetric form, is that the unification occurs
at an energy scale which is enormously large by the standards of
traditional particle physics, perhaps approaching $10^{16-17}$ Gev.
From a phenomenological viewpoint, this is fortunate.  The most
compelling unification schemes merge quarks, antiquarks, leptons, and
antileptons into common multiplets, and have gauge bosons mediating
transitions among all these particle types.  Baryon number violating
processes almost inevitably result, whose rate is inversely
proportional to the fourth power of the gauge boson masses, and thus
to the fourth power of the unification scale.  Only for such large
values of the scale is one safe from experimental limits on nucleon
instability.

From a theoretical point of view the large scale is fascinating
because it brings us, from the internal logic of particle physics, to
the threshold of quantum gravity.

I find it quite remarkable that the logarithmic running of couplings,
discovered theoretically and now amply verified within QCD, permits a
meaningful quantitative discussion of these extremely ambitious and
otherwise thinly rooted ideas, and even allows us to discriminate
between different possibilities (especially, SUSY vs. non-SUSY).

\section{Lessons: The Nature of Nature}

Since QCD is our most perfect example of a fundamental theory of
Nature, it is appropriate to use it as a basis for drawing broad
conclusions about how Nature works, or, in other words, for ``natural
philosophy''.

Let me do this by listing some adjectives we might use to describe the
theory; the implication being, that these adjectives therefore
describe Nature herself:

{\it alien} : As has been the case for all fundamental physical
theories since Galileo, QCD is formulated in abstract mathematical
terms.  In particular, there are no hints of moral concepts or
purposes.  Nor, in thousands of rigorous experiments, have we
encountered any signs of active intervention in the unfolding of the
equations according to permanent laws.

{\it simple} : In its appropriate, natural language, QCD can be
written in one line, using only symbols that cleanly embody its
conceptual basis.

{\it beautiful} : That she achieves so much with such economy of
means, marks Nature as a skillful artist.  She plays with symmetries,
creating and destroying them in varied, fascinating ways.

{\it weird} : Quantum mechanics is notoriously weird, and QCD
incorporates it in its marrow.  Less remarkable, but to me no less
weird, is the need to define QCD through a limiting procedure.  This
would seem to be a rather difficult and inefficient way to run a
Universe.

{\it comprehensible} : QCD wonderfully illustrates Einstein's
remark, ``The most incomprehensible thing about Nature is that it is
comprehensible.''  One hundred years ago people did not know there
were such things as an atomic nuclei and a strong interaction; just
over fifty years ago the pion and kaon were discovered; just over
twenty-five years ago the strong interaction problem still seemed
hopelessly intractable.  That we, collectively, have got from there to
here so quickly, against overwhelming odds, is an extraordinary
achievement.  It is a tribute to our culture, and to the glory of the
human mind.  And it is there that where we must locate, for now, the
most incomprehensible thing about Nature.

\begin{description}

\item{\bf Figure Captions}

\item{\bf Figure 1}
Experimental verification of the running of the coupling, as
predicted in \cite{four}.  The determinations, running from
left to right, are from: corrections to Bjorken sum rule, corrections
to Gross-Llewyllen-Smith sum rule, hadronic width of $\tau$ lepton, b
$\bar {\rm b}$ threshold production, prompt photon production in pp and
$\bar{\rm p}$ collisions, scaling violation in deeply ineslastic scattering,
lattice gauge theory calculations for heavy quark spectra, heavy
quarkonium decays, shape variables characterizing jets at different
energies (white dots), total $e^+e^-$ annihilation cross section, jet
production in semileptonic and hadronic processes, energy dependence
of photons in Z decay, W production, and electroweak radiative
corrections.  From \cite{fifteen}.

\item{\bf Figure 2}
Evolution of the structure function F$_2$, as measured and compared
with QCD predictions (solid lines).  From \cite{sixteen}.

\item{\bf Figure 3}
Energy and angular characterization of three-jet events,
testing the basic quark-gluon vertex very directly.  From
\cite{eighteen}.

\item{\bf Figure 4}
Comparison of the hadronic spectrum with first-principles
calculations from QCD, using techniques of lattice gauge theory.  From
\cite{nineteen}.

\item{\bf Figure 5}
Top part: Evolution of the energy and of the pressure of
2-flavor QCD as a function of temperature, showing precocious and
rapid approach to a quasi-free quark-gluon plasma.  Bottom part:
Evolution of the chiral condensation order parameter and of the
Polyakov loop, which is a measure of the inverse induced mass of an
inserted color source, and vanishes in a confined phase.  One sees
clear signals of chiral symmetry restoration and deconfinement
\cite{twentyone}.

\item{\bf Figure 6}
Connecting the second- and first- order chiral symmetry
restoration transitions predicted for two, respectively three, light
quark flavors.  The end of the first-order line is a tricritical
point.  A similar diagram may be valid for fixed strange quark mass
and varying chemical potential.

\item{\bf Figure 7}
Running of the couplings extrapolated toward very high
scales, using just the fields of the standard model.  The couplings do
not quite meet.  Experimental uncertainties in the extrapolation are
indicated by the width of the lines \cite{thirtyfour}.

\item{\bf Figure 8}
Running of the couplings extrapolated to high scales,
including the effects of supersymmetric particles starting at 1 Tev.
Within experimental and theoretical uncertainties, the couplings do
meet \cite{thirtyfive}.

\end{description}


\newpage

\begin{thebibliography}{9}

\bibitem{one}  F. Wilczek, Rev. Mod. Phys. 71, S85-S95 (1999).

\bibitem{two}  K. Wilson, Phys. Rev. D10, 2445 (1975).

\bibitem{three}  S. Coleman and  D. Gross, Phys. Rev. Lett. 31, 851 (1973).


\bibitem{four}  D. Gross and F. Wilczek, Phys. Rev. Lett. 30, 1343 (1973);
H. Politzer, Phys. Rev. Lett. 30, 1346 (1973).

\bibitem{five} D. Gross and F. Wilczek, Phys. Rev. D8, 3633 (1973),
Phys. Rev. D9, 980 (1974); H. Georgi and H. Politzer, Phys. Rev. D9,
416 (1974).

\bibitem{six} Reviewed in H. Leutwyler, hep-ph/9609465.

\bibitem{seven} T. Banks, S. Raby, L. Susskind, J. Kogut, D. Jones,
P. Scharbach, and D. Sinclair, Phys. Rev. 15, 1111 (1977). 

\bibitem{eight} M. Creutz, Phys. Rev. D21, 2308 (1980). 

\bibitem{nine} S. Coleman and E. Weinberg, Phys. Rev. D7, 1888 (1973).

\bibitem{ten}  R. Peccei and H. Quinn, Phys. Rev. D16, 1791 (1977).

\bibitem{eleven} S. Weinberg, Phys. Rev. Lett. 40, 223 (1978); F. Wilczek,
Phys. Rev. Lett. 40, 279 (1978).

\bibitem{twelve}  F. Wilczek, Nature, 397, 303-306 (1999).

\bibitem{thirteen}  S. Aoki et al., hep-lat/9904012.

\bibitem{fourteen} J. Gasser, H. Leutwyler and M. Sainio,
    Phys. Lett. B253, 252, 260 (1991).

\bibitem{fifteen} M. Schmelling, preprint MPI-H-V39, hep-ex/9701002.
Talk given at the 28th International Conference on High-energy Physics
(ICHEP96), Warsaw, Poland (1996).

\bibitem{sixteen} R.D. Ball, hep-ph/9609309.  Summary Talk of WG1, in
proceedings of DIS96, Rome, April 1996.

\bibitem{seventeen}  A. de Rujula, S. Glashow, H. Politzer,
   S. Treiman, F. Wilczek and A. Zee, Phys. Rev. D10, 2216 (1974).

\bibitem{eighteen} S. Bethke and J.E. Pilcher,
                 Ann. Rev. Nucl. Part. Sci., 42, 251-289, (1992)

\bibitem{nineteen} R. Burkhalter, http://xxx.lanl.gov/abs/hep-lat/9810043.

\bibitem{twenty}  D. Kaplan, Phys. Lett. B288, 342 (1992), Nucl. Phys. B30
 (Proc. Suppl.) 597 (1993).


\bibitem{twentyone} S. Gottlieb, et al., Phys. Rev. D47, 3619 (1993).

\bibitem{twentytwo} J. Andersen, E. Braaten, and M. Strickland,
   hep-ph/9905337; J.-P. Blaizot, E. Iancu and A. Rebhan, hep-ph/9906340.
 

\bibitem{twentythree}  M. Stephanov, K. Rajagopal and E. Shuryak,
Phys. Rev. Lett. 81, 4816 (1998).  hep-ph/9806219.

\bibitem{twentyfour} R. Pisarski and F. Wilczek, Phys. Rev. D29, (1984).

\bibitem{twentyfive} F. Wilczek, Int. J. Mod. Phys A7 3911 (1992).  


\bibitem{twentysix}  M. Stephanov, K. Rajagopal and E. Shuryak,
hep-ph/9903292.

\bibitem{twentyseven} M. Alford, A. Kapustin and F. Wilczek,
Phys. Rev. D59, 054502 (1999).

\bibitem{twentyeight} M. Alford, K. Rajagopal and F. Wilczek,
Nucl. Phys. B537, 443-458 (1999). 

\bibitem{twentynine}  M.  Han and Y. Nambu, Phys. Rev. 139B, 1006 (1965).

\bibitem{thirty} J. Sakurai, Annals of Physics 11, 1 (1960). 

\bibitem{thirtyone} C. N. Yang and R. Mills, Phys. Rev. 96, 191 (1954). 

\bibitem{hadron} T. Schaefer and F. Wilczek, Accepted in
Phys. Rev. Lett., 82, (1999)


\bibitem{thirtytwo} F. Wilczek, The Recent Excitement in High-Density
QCD, invited talk at PANIC `99, Uppsala, Sweden, June 10,
1999. IASSNS-HEP/99-66.  Preprint in preparation.  

\bibitem{thirtythree} For a short and a long review of the
subject, see respectively  S. Dimopoulos, S. Raby, and
F. Wilczek,Physics Today 44, 25 (1991); K. Dienes, Physics Reports 287,
447-525 (1997) 

\bibitem{thirtyfour} P. Langacker and M. Luo, Phys. Rev. D44, 817-822 (1991).

\bibitem{thirtyfive}  Figure 8 courtesy of K. Dienes, CERN Theory Group.


\end{thebibliography}
\end{document}